\documentclass[a4paper]{report}
\usepackage[utf8]{inputenc}
\usepackage[T1]{fontenc}
\usepackage{RJournal}
\usepackage{amsmath,amssymb,array}
\usepackage{booktabs}

\providecommand{\tightlist}{%
  \setlength{\itemsep}{0pt}\setlength{\parskip}{0pt}}

\begin{document}

\sectionhead{Research article}
\volume{}
\volnumber{}
\year{2024}
\month{}

\begin{article}
  \title{Jaya R Package - A Parameter-Free Solution for Advanced Single
  	and Multi-Objective Optimization}
  \author{by Neeraj Dhanraj Bokde}
  
  \maketitle
  
  \abstract{%
  	The \texttt{Jaya} R package offers a robust and versatile implementation
  	of the parameter-free Jaya optimization algorithm, suitable for solving
  	both single-objective and multi-objective optimization problems. By
  	integrating advanced features such as constraint handling, adaptive
  	population management, Pareto front tracking for multi-objective
  	trade-offs, and parallel processing for computational efficiency, the
  	package caters to a wide range of optimization challenges. Its intuitive
  	design and flexibility allow users to solve complex, real-world problems
  	across various domains. To demonstrate its practical utility, a case
  	study on energy modeling explores the optimization of renewable energy
  	shares, showcasing the package's ability to minimize carbon emissions
  	and costs while enhancing system reliability. The \texttt{Jaya} R
  	package is an invaluable tool for researchers and practitioners seeking
  	efficient and adaptive optimization solutions.
  }
  
  \hypertarget{introduction}{%
  	\subsection{Introduction}\label{introduction}}
  
  Optimization is a fundamental process in solving complex problems across
  diverse domains, including engineering, finance, healthcare, and energy
  systems. Traditional optimization methods often rely on gradient
  information, which limits their applicability to non-smooth,
  multi-modal, or highly constrained problems. To address these
  challenges, gradient-free optimization algorithms like the Jaya
  algorithm have emerged as powerful alternatives.
  
  The Jaya algorithm \citep{rao2016jaya} is a population-based,
  parameter-free optimization technique. It minimizes or maximizes
  objective functions by iteratively refining a set of candidate solutions
  without requiring hyperparameters. Its simplicity, adaptability, and
  robustness make it a suitable choice for single-objective and
  multi-objective problems, even under complex constraints.
  
  The \textbf{Jaya R package} \citep{Jaya} provides an efficient and
  flexible implementation of the Jaya algorithm within the R ecosystem. It
  supports both single-objective and multi-objective optimization,
  integrating advanced features such as: Adaptive population adjustment to
  improve convergence rates, Constraint handling for practical
  feasibility, and, Pareto front tracking for multi-objective problems.
  
  Designed to address the needs of researchers and practitioners, the
  package is equipped with tools for early stopping, parallel processing,
  and seamless visualization of results. It leverages R's extensive
  ecosystem to offer an intuitive yet powerful optimization framework.
  
  This paper introduces the Jaya R package, emphasizing its design,
  features, and practical utility. While the package is applicable across
  a wide range of disciplines, we present an example use case in energy
  modeling to demonstrate its capabilities. The case study showcases how
  the package can optimize renewable energy shares to achieve objectives
  such as minimizing carbon emissions and costs while maximizing system
  reliability. However, the focus of this paper remains on the Jaya R
  package itself, with the case study serving as a practical illustration.
  
  Subsequent sections provide a detailed overview of the package's
  features (Section 2), demonstrate its use in energy modeling (Section
  3), and conclude with a discussion on its advantages and potential
  applications (Section 4).
  
  \hypertarget{the-jaya-algorithm}{%
  	\subsubsection{The Jaya Algorithm}\label{the-jaya-algorithm}}
  
  The Jaya algorithm, introduced in \citep{rao2016jaya}, is a
  population-based optimization technique designed for solving
  single-objective and multi-objective problems. The term ``Jaya''
  translates to ``victory'' in Sanskrit, reflecting the algorithm's
  philosophy of moving toward the best solution while avoiding the worst.
  Unlike traditional optimization algorithms that often require
  algorithm-specific parameters, such as crossover rates or mutation
  probabilities, the Jaya algorithm is parameter-free. This simplicity,
  coupled with its effectiveness, makes it an attractive choice for
  optimization across diverse domains.
  
  The Jaya algorithm begins by generating a population of candidate
  solutions, each representing a possible configuration of decision
  variables within their specified bounds. Once initialized, the algorithm
  evaluates each candidate solution using one or more objective functions.
  For constrained problems, penalty terms are applied to account for
  constraint violations, ensuring that the algorithm prioritizes feasible
  solutions.
  
  The core of the algorithm lies in its update rule. Each candidate
  solution is adjusted based on its proximity to the best and worst
  solutions in the current population. This adjustment is governed by the
  following equation:
  
  \[
  x_{\text{new}} = x_{\text{old}} + r_1 \cdot (x_{\text{best}} - |x_{\text{old}}|) - r_2 \cdot (x_{\text{worst}} - |x_{\text{old}}|)
  \]
  
  Here, \(x_{\text{new}}\) represents the updated solution,
  \(x_{\text{old}}\) is the current solution, \(x_{\text{best}}\) is the
  best solution in the population, and \(x_{\text{worst}}\) is the worst
  solution. The terms \(r_1\) and \(r_2\) are random numbers uniformly
  distributed between 0 and 1, introducing stochasticity that aids in
  exploring the solution space. This simple update mechanism enables the
  algorithm to refine solutions iteratively.
  
  The algorithm repeats the evaluation and update steps until a specified
  number of iterations is reached or a convergence criterion is met. The
  output of the Jaya algorithm is the best solution(s) found, along with
  their corresponding objective values. The absence of algorithm-specific
  parameters eliminates the need for manual tuning, further enhancing its
  usability.
  
  Figure \ref{fig:block} is a block diagram that illustrates the workflow
  of the Jaya algorithm, encompassing initialization, evaluation, and
  iterative improvement:
  
  \begin{Schunk}
  	\begin{figure}
  		
  		{\centering \includegraphics[width=0.9\linewidth]{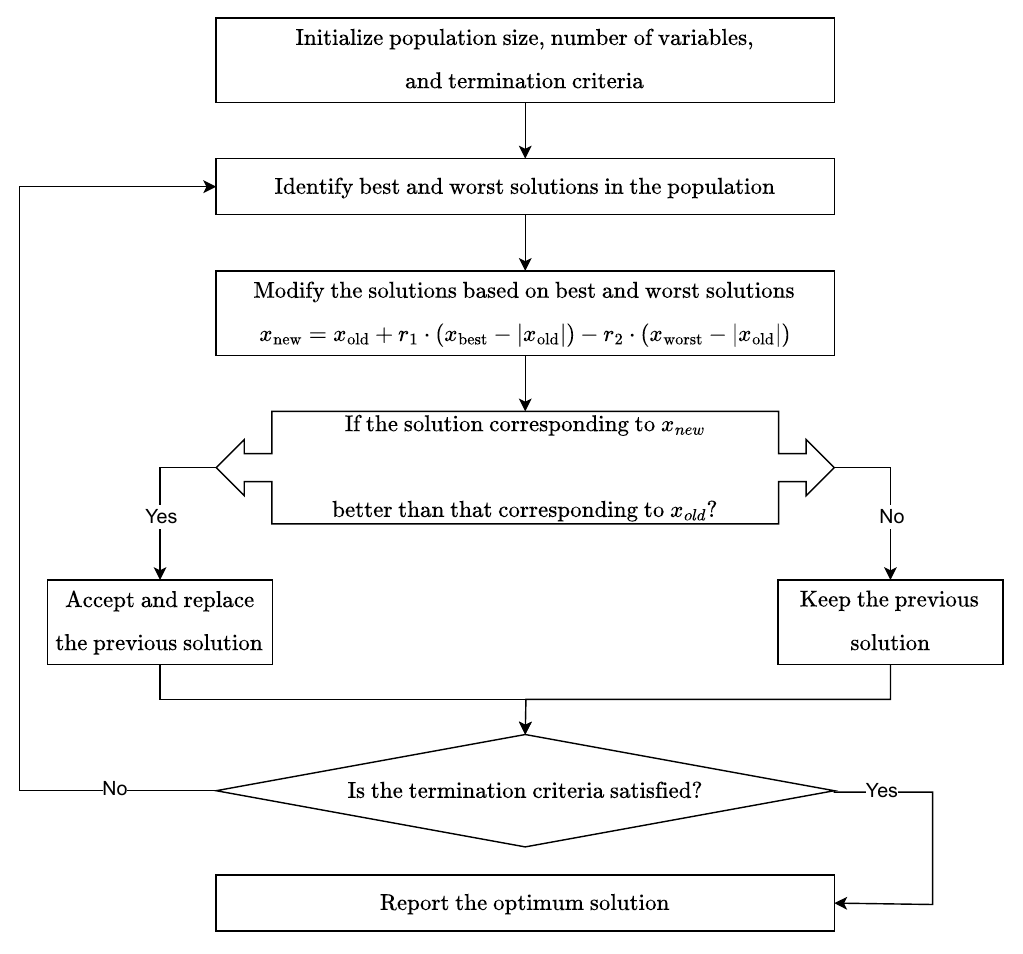} 
  			
  		}
  		
  		\caption[Block diagram of the Jaya algorithm]{Block diagram of the Jaya algorithm.}\label{fig:block}
  	\end{figure}
  \end{Schunk}
  
  The algorithm's simplicity, versatility, and parameter-free nature make
  it particularly well-suited for applications in engineering, energy
  modeling, and other fields requiring robust optimization techniques. The
  Jaya R package builds upon these foundational principles to provide a
  user-friendly interface for implementing this powerful algorithm within
  the R programming environment.
  
  \hypertarget{overview-of-the-jaya-r-package}{%
  	\subsection{Overview of the Jaya R
  		Package}\label{overview-of-the-jaya-r-package}}
  
  The \textbf{Jaya R package} offers a comprehensive implementation of the
  Jaya optimization algorithm, a gradient-free, parameter-less technique
  suitable for solving single-objective and multi-objective optimization
  problems. Its design focuses on flexibility, simplicity, and
  applicability across diverse domains.
  
  \hypertarget{key-features-of-the-jaya-r-package}{%
  	\subsubsection{\texorpdfstring{Key Features of the \texttt{Jaya} R
  			Package}{Key Features of the Jaya R Package}}\label{key-features-of-the-jaya-r-package}}
  
  The \texttt{Jaya} R package incorporates several advanced features that
  enhance its applicability for solving both single-objective and
  multi-objective optimization problems. These features are designed to
  provide users with flexibility, ease of use, and computational
  efficiency. The key features are as follows:
  
  \begin{enumerate}
  	\def\labelenumi{\arabic{enumi}.}
  	\item
  	\textbf{Single-Objective Optimization:}\\
  	The package supports robust optimization for single objectives, with
  	capabilities to define custom objective functions, parameter bounds,
  	and constraints.
  	\item
  	\textbf{Multi-Objective Optimization:}\\
  	The implementation of Pareto front tracking allows users to
  	efficiently solve multi-objective problems, balancing trade-offs
  	across conflicting objectives.
  	\item
  	\textbf{Constraint Handling:}\\
  	User-defined constraint functions are supported, enabling the
  	algorithm to prioritize feasible solutions by enforcing restrictions
  	through penalty terms.
  	\item
  	\textbf{Adaptive Population Adjustment:}\\
  	The population size is dynamically adjusted during the optimization
  	process, enhancing convergence by balancing exploration and
  	exploitation phases.
  	\item
  	\textbf{Early Stopping Mechanism:}\\
  	The algorithm incorporates early termination criteria based on
  	user-defined tolerance and patience settings, improving runtime
  	efficiency for problems nearing convergence.
  	\item
  	\textbf{Parallel Processing:}\\
  	The package leverages multi-core systems to parallelize objective
  	function evaluations, significantly reducing computation time for
  	large-scale problems.
  \end{enumerate}
  
  These features make the \texttt{Jaya} R package a versatile and powerful
  tool for optimization in diverse domains such as engineering, energy
  modeling, and operational research.
  
  \hypertarget{primary-functions-in-the-package}{%
  	\subsubsection{Primary Functions in the
  		Package}\label{primary-functions-in-the-package}}
  
  \hypertarget{jaya-single-objective-optimization}{%
  	\paragraph{1. jaya(): Single-Objective
  		Optimization}\label{jaya-single-objective-optimization}}
  
  The jaya() function solves optimization problems with a single objective
  function. It allows constraints and offers options for adaptive
  population size adjustment.
  
  \textbf{Usage}:
  
  \begin{verbatim}
  	jaya(fun, lower, upper, popSize = 50, maxiter, n_var, seed = NULL,
  	constraints = list(), adaptive_pop = FALSE, ...)
  \end{verbatim}
  
  \textbf{Parameters}:
  
  \begin{itemize}
  	\item
  	fun: The objective function to minimize or maximize.
  	\item
  	lower: A numeric vector of lower bounds for decision variables.
  	\item
  	upper: A numeric vector of upper bounds for decision variables.
  	\item
  	popSize: The size of the population (default = 50).
  	\item
  	maxiter: The maximum number of iterations.
  	\item
  	n\_var: The number of decision variables.
  	\item
  	seed: An optional seed for reproducibility.
  	\item
  	constraints: A list of constraint functions that return values \(\le\)
  	0 for feasibility.
  	\item
  	adaptive\_pop: Logical; enables dynamic population size adjustment
  	(default = FALSE).
  \end{itemize}
  
  \textbf{Output}: The function returns a list containing:
  
  \begin{itemize}
  	\tightlist
  	\item
  	Best: The best solution found, including decision variables and the
  	objective value.
  	\item
  	Iterations: A history of the best objective values over iterations.
  \end{itemize}
  
  \hypertarget{jaya_multi-multi-objective-optimization}{%
  	\paragraph{2. jaya\_multi(): Multi-Objective
  		Optimization}\label{jaya_multi-multi-objective-optimization}}
  
  The jaya\_multi() function extends the algorithm for problems with
  multiple objectives. It identifies Pareto-optimal solutions and supports
  constraints.
  
  \textbf{Usage}:
  
  \begin{verbatim}
  	jaya_multi(objectives, lower, upper, popSize = 50, maxiter, n_var,
  	seed = NULL, constraints = list(), adaptive_pop = FALSE, ...)
  \end{verbatim}
  
  \textbf{Parameters}:
  
  \begin{itemize}
  	\tightlist
  	\item
  	objectives: A list of objective functions.
  	\item
  	lower: A numeric vector of lower bounds for decision variables.
  	\item
  	upper: A numeric vector of upper bounds for decision variables.
  	\item
  	popSize: The size of the population (default = 50).
  	\item
  	maxiter: The maximum number of iterations.
  	\item
  	n\_var: The number of decision variables.
  	\item
  	seed: An optional seed for reproducibility.
  	\item
  	constraints: A list of constraint functions that return values \(\le\)
  	0 for feasibility.
  	\item
  	adaptive\_pop: Logical; enables dynamic population size adjustment
  	(default = FALSE).
  \end{itemize}
  
  \textbf{Output}:
  
  The function returns:
  
  \begin{itemize}
  	\tightlist
  	\item
  	Pareto\_Front: A data frame of non-dominated solutions, including
  	decision variables and their corresponding objective values.
  	\item
  	Solutions: The final population, including all solutions evaluated.
  \end{itemize}
  
  \hypertarget{helper-functions}{%
  	\subsubsection{Helper Functions}\label{helper-functions}}
  
  \hypertarget{summary.jaya}{%
  	\paragraph{\texorpdfstring{\texttt{summary.jaya()}}{summary.jaya()}}\label{summary.jaya}}
  
  The \texttt{summary.jaya()} function provides a concise summary of the
  optimization results for both single-objective (\texttt{jaya()}) and
  multi-objective (\texttt{jaya\_multi()}) optimization runs. It includes:
  
  \begin{itemize}
  	\tightlist
  	\item
  	The \textbf{best solution} found during the optimization process.
  	\item
  	The \textbf{objective value} of the best solution.
  	\item
  	Convergence details, such as the number of iterations taken to reach
  	the optimal solution.
  \end{itemize}
  
  This function is particularly useful for quickly understanding the
  results of the optimization process.
  
  \textbf{Usage}:
  
  \begin{verbatim}
  	summary(object)
  \end{verbatim}
  
  \textbf{Arguments}:
  
  \begin{itemize}
  	\tightlist
  	\item
  	\texttt{object}: An object of class \texttt{jaya} or
  	\texttt{jaya\_multi}, which is the output of the \texttt{jaya()} or
  	\texttt{jaya\_multi()} functions.
  \end{itemize}
  
  \hypertarget{plot.jaya}{%
  	\paragraph{\texorpdfstring{\texttt{plot.jaya()}}{plot.jaya()}}\label{plot.jaya}}
  
  The \texttt{plot.jaya()} function visualizes the progress of the
  optimization for single-objective problems. It generates a plot of the
  \textbf{best objective value} versus the \textbf{number of iterations},
  allowing users to assess convergence behavior.
  
  \textbf{Features}:
  
  \begin{itemize}
  	\tightlist
  	\item
  	Easy identification of convergence trends.
  	\item
  	Highlights stagnation or rapid improvement phases during optimization.
  \end{itemize}
  
  \textbf{Usage}:
  
  \begin{verbatim}
  	plot(object)
  \end{verbatim}
  
  \textbf{Arguments}:
  
  \begin{itemize}
  	\tightlist
  	\item
  	\texttt{object}: An object of class \texttt{jaya}, which is the output
  	of the \texttt{jaya()} function.
  \end{itemize}
  
  \textbf{Output}:
  
  \begin{itemize}
  	\tightlist
  	\item
  	A convergence plot displaying the evolution of the best objective
  	value over iterations.
  \end{itemize}
  
  \hypertarget{plot_jaya_multi_pairwise}{%
  	\paragraph{\texorpdfstring{\texttt{plot\_jaya\_multi\_pairwise()}}{plot\_jaya\_multi\_pairwise()}}\label{plot_jaya_multi_pairwise}}
  
  The \texttt{plot\_jaya\_multi\_pairwise()} function generates pairwise
  scatter plots for the objectives in the Pareto front, aiding in the
  analysis of trade-offs between conflicting objectives for
  multi-objective optimization.
  
  \textbf{Features}:
  
  \begin{itemize}
  	\tightlist
  	\item
  	Visualization of the Pareto front in 2D projections.
  	\item
  	Identifies clusters or patterns in trade-offs between objectives.
  \end{itemize}
  
  \textbf{Usage}:
  
  \begin{verbatim}
  	plot_jaya_multi_pairwise(object, ... )
  \end{verbatim}
  
  \textbf{Arguments}:
  
  \begin{itemize}
  	\tightlist
  	\item
  	\texttt{object}: An object of class \texttt{jaya\_multi}, which is the
  	output of the \texttt{jaya\_multi()} function.
  	\item
  	Additional arguments for customization of plot aesthetics.
  \end{itemize}
  
  \textbf{Output}:
  
  \begin{itemize}
  	\tightlist
  	\item
  	Pairwise scatter plots of objectives from the Pareto front.
  \end{itemize}
  
  \hypertarget{functionality-overview}{%
  	\subsection{Functionality Overview}\label{functionality-overview}}
  
  \hypertarget{single-objective-optimization}{%
  	\subsubsection{Single-Objective
  		Optimization}\label{single-objective-optimization}}
  
  The \texttt{jaya()} function supports optimization problems involving a
  single objective function. Users can define the objective function to
  minimize or maximize, specify bounds for the decision variables, and
  configure additional parameters such as population size and maximum
  iterations. The function iteratively refines the solutions until
  convergence criteria are met, ultimately providing the best solution
  found along with its corresponding objective value. Furthermore, it
  tracks the evolution of the best objective value over iterations,
  enabling users to visualize the optimization process.
  
  In the following example, we optimize a simple convex function known as
  the Sphere function, which is defined as the sum of the squares of the
  input vector. This function is commonly used as a benchmark in
  optimization due to its simplicity and unimodal nature.
  
  \begin{Schunk}
  	\begin{Sinput}
  		library(Jaya)  # Load the Jaya optimization package
  		
  		# Define the Sphere function
  		# The Sphere function calculates the sum of squares of the input vector x.
  		# It is a simple convex function commonly used for testing optimization algorithms.
  		sphere_function <- function(x) sum(x^2)
  		
  		# Perform single-objective optimization using the Jaya algorithm
  		result <- jaya(
  		fun = sphere_function,         # Objective function to minimize
  		lower = c(-5, -5, -5),         # Lower bounds for the variables
  		upper = c(5, 5, 5),            # Upper bounds for the variables
  		popSize = 20,                  # Population size
  		maxiter = 50,                  # Maximum number of iterations
  		n_var = 3                      # Number of variables
  		)
  		
  		# Display a summary of the optimization results
  		summary(result)
  	\end{Sinput}
  	\begin{Soutput}
  		#> Jaya Algorithm
  		#> Population Size      = 20 
  		#> Number of iterations = 50 
  		#> Number of variables  = 3 
  		#> 
  		#> Objective: minimize 
  		#> Objective Function:
  		#> [[1]]
  		#> function (x) 
  		#> sum(x^2)
  		#> <bytecode: 0x000001a912032a88>
  		#> 
  		#> 
  		#> Limits:
  		#> x1 = [-5, 5]
  		#> x2 = [-5, 5]
  		#> x3 = [-5, 5]
  		#> 
  		#> Best Result:
  		#>            Best.x1      Best.x2      Best.x3    Best.f.x.
  		#> Best -6.373456e-05 5.722159e-05 9.972234e-05 1.728095e-08
  	\end{Soutput}
  	\begin{Sinput}
  		# Plot the convergence of the best objective value over iterations
  		plot(result)
  	\end{Sinput}
  	\begin{figure}[htbp]
  		
  		{\centering \includegraphics[width=5in]{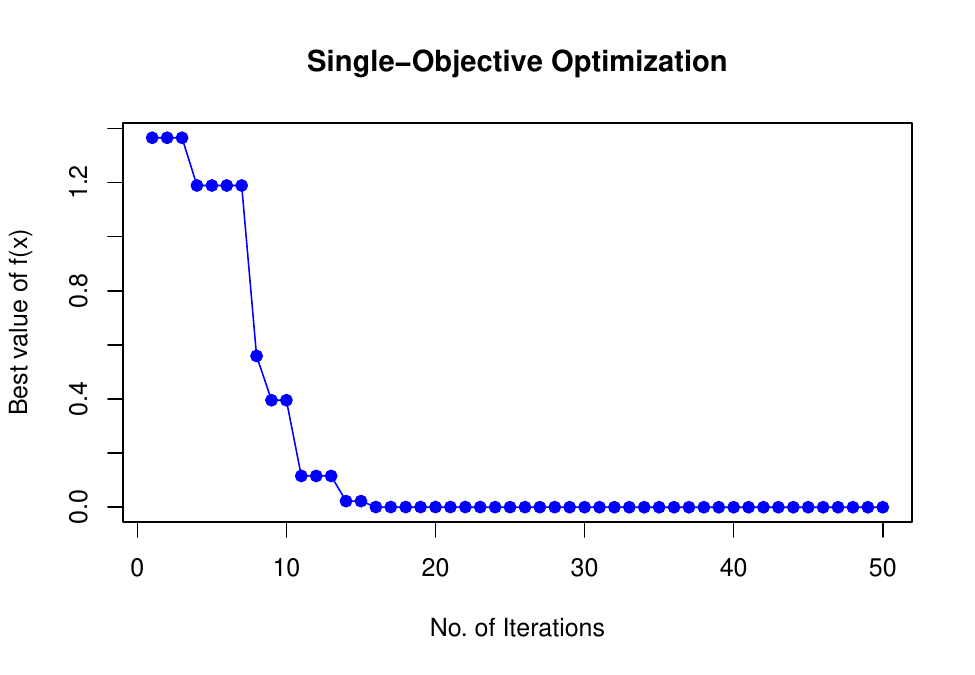} 
  			
  		}
  		
  		\caption[Convergence plot showing the evolution of the best objective value over iterations for minimizing a sphere function using the Jaya algorithm]{Convergence plot showing the evolution of the best objective value over iterations for minimizing a sphere function using the Jaya algorithm.}\label{fig:opt1}
  	\end{figure}
  \end{Schunk}
  
  \begin{itemize}
  	\item
  	\textbf{Objective Function}: The Sphere function,
  	\(f(x) = \sum_{i=1}^n x_i^2\), is used as the objective function. It
  	evaluates how close the variables are to zero, with the global minimum
  	at \(f(x) = 0\), when \(x_i = 0\) for all \(i\).
  	\item
  	\textbf{Bounds}: Decision variables are bounded between -5 and 5,
  	ensuring the search is confined within this range.
  	\item
  	\textbf{Parameters}:
  	
  	\begin{itemize}
  		\tightlist
  		\item
  		\texttt{popSize\ =\ 20}: Specifies a population of 20 candidate
  		solutions.
  		\item
  		\texttt{maxiter\ =\ 50}: Runs the algorithm for a maximum of 50
  		iterations.
  		\item
  		\texttt{n\_var\ =\ 3}: Optimizes three decision variables.
  	\end{itemize}
  \end{itemize}
  
  The output of the \texttt{summary(result)} command provides a concise
  overview of the optimization process, including the best solution found,
  the corresponding objective value, and convergence details.
  
  The \texttt{plot(result)} function generates a convergence plot (Figure
  \ref{fig:opt1}) showing the evolution of the best objective value over
  iterations. This plot helps assess the efficiency of the algorithm in
  minimizing the objective function and provides insights into whether the
  optimization process has stagnated or continued to improve.
  
  \hypertarget{multi-objective-optimization}{%
  	\subsubsection{Multi-Objective
  		Optimization}\label{multi-objective-optimization}}
  
  The \texttt{jaya\_multi()} function facilitates the optimization of
  multiple conflicting objectives by employing Pareto front tracking. This
  approach identifies non-dominated solutions that represent trade-offs
  among the objectives, enabling users to explore the solution space
  without biasing toward a single objective. The function accepts a list
  of objective functions, bounds for the decision variables, and
  additional parameters to control the optimization process. The result
  includes the Pareto-optimal solutions and their corresponding decision
  variables, which are critical for understanding trade-offs in
  multi-objective optimization problems.
  
  In the following example, we optimize two objectives simultaneously:
  
  \begin{itemize}
  	\tightlist
  	\item
  	Minimize the sum of squares of decision variables (Sphere function).
  	\item
  	Minimize the sum of squares offset by 2.
  \end{itemize}
  
  This example highlights how the Jaya algorithm balances competing
  objectives to generate a Pareto front.
  
  \begin{Schunk}
  	\begin{Sinput}
  		# Multi-objective example: Optimize two sphere functions
  		objective1 <- function(x) sum(x^2)                      # Minimize the sum of squares
  		objective2 <- function(x) sum((x - 2)^2)                # Minimize the sum of squares offset by 2
  		
  		# Run the multi-objective optimization
  		result_multi <- jaya_multi(
  		objectives = list(objective1, objective2),            # Add the objective functions
  		lower = c(-5, -5, -5),                                # Lower bounds for variables
  		upper = c(5, 5, 5),                                   # Upper bounds for variables
  		popSize = 30,                                         # Population size
  		maxiter = 100,                                        # Maximum number of iterations
  		n_var = 3                                             # Number of variables
  		)
  		
  		# Summarize the results
  		summary(result_multi)
  	\end{Sinput}
  	\begin{Soutput}
  		#>              Length Class      Mode
  		#> Pareto_Front 5      data.frame list
  		#> Solutions    5      data.frame list
  	\end{Soutput}
  	\begin{Sinput}
  		# Plot pairwise scatter plots of the Pareto front
  		plot_jaya_multi_pairwise(result_multi)
  	\end{Sinput}
  	\begin{figure}[htbp]
  		
  		{\centering \includegraphics[width=5in]{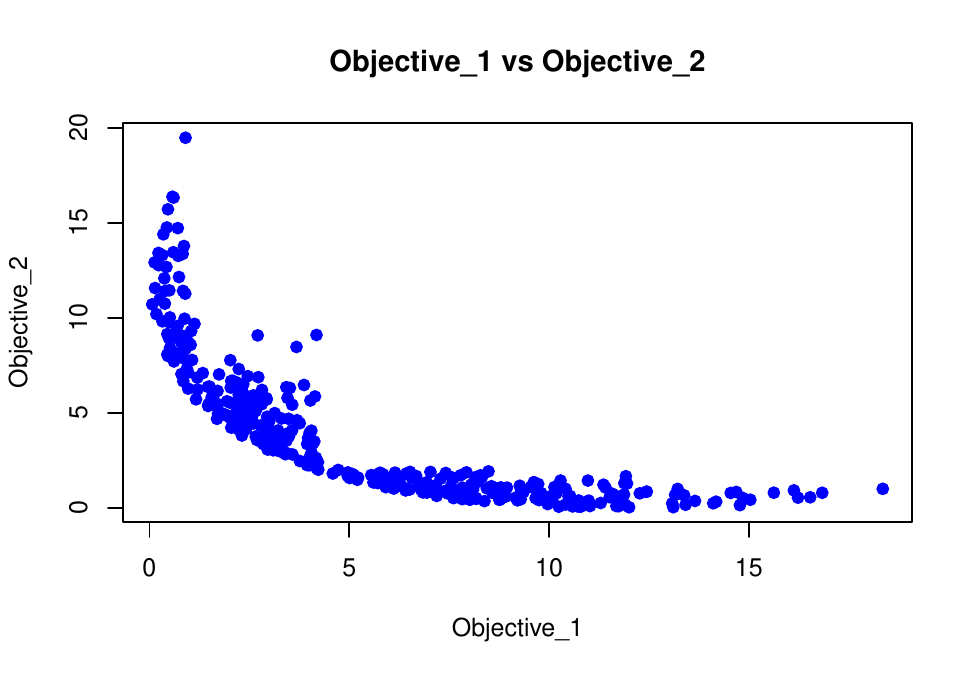} 
  			
  		}
  		
  		\caption[Pairwise scatter plots of the Pareto front obtained from the multi-objective optimization of two sphere functions]{Pairwise scatter plots of the Pareto front obtained from the multi-objective optimization of two sphere functions. The plots show trade-offs between objectives: minimizing the sum of squares and minimizing the sum of squares offset by 2.}\label{fig:opt2}
  	\end{figure}
  \end{Schunk}
  
  \begin{itemize}
  	\item
  	\textbf{Objective Functions}:
  	
  	\begin{itemize}
  		\tightlist
  		\item
  		\texttt{objective1(x)}: The Sphere function
  		\(f(x) = \sum_{i=1}^n x_i^2\), a convex function, is minimized to
  		push variables close to zero.
  		\item
  		\texttt{objective2(x)}: A modified Sphere function
  		\(f(x) = \sum_{i=1}^n (x_i - 2)^2\), aims to shift variables towards
  		2.
  	\end{itemize}
  	\item
  	\textbf{Parameters}:
  	
  	\begin{itemize}
  		\tightlist
  		\item
  		\texttt{popSize\ =\ 30}: A population size of 30 candidate solutions
  		is used.
  		\item
  		\texttt{maxiter\ =\ 100}: The algorithm runs for a maximum of 100
  		iterations.
  		\item
  		\texttt{n\_var\ =\ 3}: The problem involves three decision
  		variables.
  		\item
  		\texttt{lower} and \texttt{upper}: Decision variables are
  		constrained to lie within {[}-5, 5{]}.
  	\end{itemize}
  \end{itemize}
  
  The \texttt{summary(result\_multi)} command provides a detailed overview
  of the Pareto-optimal solutions, including their decision variables and
  corresponding objective values. This helps in analyzing the trade-offs
  between the objectives.
  
  The \texttt{plot\_jaya\_multi\_pairwise(result\_multi)} function
  generates pairwise scatter plots (Figure \ref{fig:opt2}) of the Pareto
  front. Each plot visualizes trade-offs between two objectives, aiding in
  identifying clusters or patterns in the solution space. In this example,
  the scatter plots reveal the relationship between minimizing the sum of
  squares and minimizing the sum of squares offset by 2.
  
  \hypertarget{performance-comparison-of-optimization-algorithms}{%
  	\subsubsection{Performance Comparison of Optimization
  		Algorithms}\label{performance-comparison-of-optimization-algorithms}}
  
  To assess the performance of the Jaya algorithm, we conducted a
  comparative study using five well-known single-objective optimization
  problems: Sphere, Rastrigin, Rosenbrock, Ackley, and Griewank. These
  problems are widely used benchmarks, offering a variety of challenges
  such as non-linearity, multimodality, and a rugged objective space. The
  performance of Jaya was compared with other optimization techniques,
  including Differential Evolution (DE)
  \citep[\citet{ardia2011differential}]{mullen2011deoptim}, Genetic
  Algorithm (GA) \citep[\citet{RJ-2017-008}]{scrucca2013ga}, Simulated
  Annealing (SA) \citep{stats}, Particle Swarm Optimization (PSO)
  \citep{pso}, and Nelder-Mead (NM) \citep{stats}.
  
  \hypertarget{benchmark-problems-and-results}{%
  	\paragraph{Benchmark Problems and
  		Results}\label{benchmark-problems-and-results}}
  
  The equations for the benchmark problems and the achieved values of the
  objective functions for each algorithm are presented in Table
  \ref{tab:benchmark_results}. These equations for the benchmark problems
  are based on \citep[\citet{e2}, \citet{e3}, \citet{e4}, \citet{e5}]{e1}
  and represent diverse optimization landscapes, and the results highlight
  the precision and efficiency of the Jaya algorithm in comparison to
  other methods. The code for this comparison study is publicly available
  at \citep{dhanraj2024jaya}.
  
  \begin{table*}[ht]
  	\footnotesize
  	\centering
  	\caption{Achieved values for benchmark problems using different optimization algorithms.}
  	\label{tab:benchmark_results}
  	\begin{tabular}{|p{1.3cm}|p{4.2cm}|p{1cm}|p{1cm}|p{1cm}|p{1cm}|p{1cm}|p{1cm}|}
  		\hline
  		\textbf{Problem} & \textbf{Equation} & \textbf{Jaya} & \textbf{DE} & \textbf{GA} & \textbf{SA} & \textbf{PSO} & \textbf{NM} \\ \hline
  		\textbf{Sphere} \cite{e1} & $\sum_{i=1}^n x_i^2$ & $2.13 \times 10^{-101}$ & $2.74 \times 10^{-110}$ & $2.78 \times 10^{-8}$ & $0$ & $0$ & $0$ \\ \hline
  		\textbf{Rastrigin} \cite{e2} & $10n + \sum_{i=1}^n \left[x_i^2 - 10\cos(2\pi x_i)\right]$ & $0$ & $0$ & $3.33 \times 10^{-7}$ & $0$ & $0$ & $0$ \\ \hline
  		\textbf{Rosenbrock} \cite{e3} & $\sum_{i=1}^{n-1} \left[100(x_{i+1} - x_i^2)^2 + (x_i - 1)^2\right]$ & $1.74 \times 10^{-9}$ & $0$ & $7.93 \times 10^{-5}$ & $2.88 \times 10^{-2}$ & $2.96 \times 10^{-15}$ & $9.84 \times 10^{-7}$ \\ \hline
  		\textbf{Ackley} \cite{e4} & $-20\exp\left(-0.2\sqrt{\frac{1}{n}\sum_{i=1}^n x_i^2}\right) - \exp\left(\frac{1}{n}\sum_{i=1}^n \cos(2\pi x_i)\right) + 20 + e$ & $1.90 \times 10^{-11}$ & $4.44 \times 10^{-16}$ & $1.39 \times 10^{-4}$ & $4.44 \times 10^{-16}$ & $4.44 \times 10^{-16}$ & $4.44 \times 10^{-16}$ \\ \hline
  		\textbf{Griewank} \cite{e5} & $1 + \frac{1}{4000}\sum_{i=1}^n x_i^2 - \prod_{i=1}^n \cos\left(\frac{x_i}{\sqrt{i}}\right)$ & $1.45 \times 10^{-3}$ & $9.99 \times 10^{-16}$ & $6.02 \times 10^{-4}$ & $0$ & $0$ & $0$ \\ \hline
  	\end{tabular}
  \end{table*}
  
  The results illustrate the competitive performance of the Jaya algorithm
  across all benchmark problems. For simpler problems such as Sphere and
  Rastrigin, most algorithms achieved near-zero values, reflecting the
  ease of optimization. However, for complex problems like Rosenbrock and
  Ackley, Jaya demonstrated high precision, achieving values comparable to
  or better than established methods like DE and PSO.
  
  Jaya's parameter-less design provides a computationally efficient
  alternative to traditional optimization algorithms, ensuring robust
  performance across a diverse range of optimization landscapes. These
  results highlight Jaya's potential as a versatile and reliable tool for
  solving optimization problems.
  
  \hypertarget{application-of-the-jaya-package-in-multi-objective-energy-optimization}{%
  	\subsection{Application of the Jaya Package in Multi-Objective Energy
  		Optimization}\label{application-of-the-jaya-package-in-multi-objective-energy-optimization}}
  
  The complexity of modern energy systems necessitates robust optimization
  methods capable of balancing conflicting objectives. Renewable energy
  integration, in particular, introduces challenges related to
  cost-effectiveness, environmental impact, and system reliability. To
  illustrate the capability of the \textbf{Jaya} R package in solving such
  problems, we applied it to a multi-objective optimization case study,
  highlighting its adaptability and computational efficiency.
  
  \hypertarget{simulation-settings}{%
  	\subsubsection{Simulation Settings}\label{simulation-settings}}
  
  The optimization problem was designed to evaluate the contribution of
  four renewable energy sources---wind, solar, hydro, and storage---under
  conflicting objectives. These objectives were: minimizing carbon
  emissions, minimizing total costs (capital and operational), and
  maximizing reliability. The latter was quantified as a composite
  reliability index reflecting the stability of energy supply from the
  renewable sources.
  
  The decision variables, representing the percentage contribution of each
  energy source, were bounded between 10\% and 40\%, ensuring a balanced
  mix. Additionally, the total renewable contribution was constrained to
  exceed 70\%, aligning with common policy mandates for renewable energy
  deployment. The \textbf{Jaya} optimizer's parameter-less design allowed
  for seamless integration of these constraints, with the
  \texttt{jaya\_multi} function managing the multi-objective optimization
  process efficiently.
  
  The simulation settings included:
  
  \begin{itemize}
  	\tightlist
  	\item
  	Population Size: 100
  	\item
  	Maximum Iterations: 100
  	\item
  	Adaptive Population Sizing: Enabled (range: 50--200)
  	\item
  	Convergence Tolerance: \(1 \times 10^{-3}\)
  \end{itemize}
  
  The code for this simulation study is publicly available at
  \citep{dhanraj2024jaya}.
  
  \hypertarget{results-and-visualizations}{%
  	\subsubsection{Results and
  		Visualizations}\label{results-and-visualizations}}
  
  The optimization identified Pareto-optimal solutions that highlight the
  trade-offs among emissions, costs, and reliability. These solutions were
  visualized as a 3D Pareto front (Figure
  \ref{fig:fig-total-contribution}), showcasing the algorithm's ability to
  navigate complex decision spaces effectively.
  
  \hypertarget{wind-solar-and-hydro-contributions}{%
  	\paragraph{Wind, Solar, and Hydro
  		Contributions}\label{wind-solar-and-hydro-contributions}}
  
  Three additional visualizations (Figures
  \ref{fig:fig-wind-contribution}, \ref{fig:fig-solar-contribution}, and
  \ref{fig:fig-hydro-contribution}) explored the individual contributions
  of wind, solar, and hydro energy sources to the Pareto front, revealing
  their distinct roles in optimizing the energy mix:
  
  \begin{itemize}
  	\item
  	\textbf{Wind Contribution (Figure \ref{fig:fig-wind-contribution}):}
  	Higher wind contributions were primarily associated with significant
  	reductions in emissions. However, this came at the cost of increased
  	variability in reliability, reflecting the intermittency of wind
  	energy. Cost trends indicated moderate increases with higher wind
  	contributions, emphasizing the importance of complementing wind with
  	other stable sources or storage.
  	\item
  	\textbf{Solar Contribution (Figure \ref{fig:fig-solar-contribution}):}
  	Solar energy demonstrated a balanced impact on all three objectives.
  	Solutions with higher solar contributions generally achieved reduced
  	emissions while maintaining moderate costs and reliability. This
  	behavior highlights solar energy's scalability and cost-effectiveness
  	as a vital component of sustainable energy systems.
  	\item
  	\textbf{Hydro Contribution (Figure \ref{fig:fig-hydro-contribution}):}
  	Hydro energy consistently contributed to enhanced reliability, with
  	higher contributions aligning with improved system stability. However,
  	these solutions were associated with increased overall costs,
  	indicative of the capital-intensive nature of hydro energy
  	infrastructure.
  \end{itemize}
  
  \hypertarget{total-renewable-contribution}{%
  	\paragraph{Total Renewable
  		Contribution}\label{total-renewable-contribution}}
  
  A composite plot (Figure \ref{fig:fig-total-contribution}) depicting the
  total contribution from all renewable sources provided additional
  insights. Interestingly, higher total contributions were associated with
  lower reliability indices (i.e., more negative values), suggesting
  challenges in maintaining system stability with higher shares of
  intermittent sources. This underscores the need for complementary
  measures, such as advanced storage systems or grid balancing
  technologies, to mitigate these effects.
  
  \begin{Schunk}
  	\begin{figure}[htbp]
  		
  		{\centering \includegraphics[width=4in]{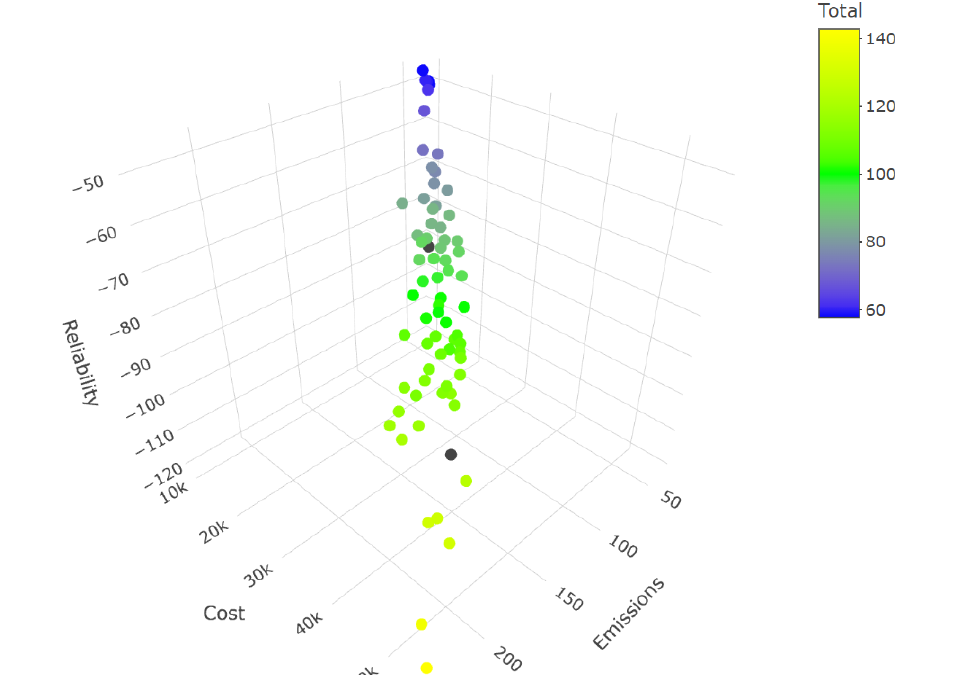} 
  			
  		}
  		
  		\caption[3D Pareto Front of Energy System Optimization (Colored by Total Contribution)]{3D Pareto Front of Energy System Optimization (Colored by Total Contribution).}\label{fig:fig-total-contribution}
  	\end{figure}
  \end{Schunk}
  
  \begin{Schunk}
  	\begin{figure}[htbp]
  		
  		{\centering \includegraphics[width=4in]{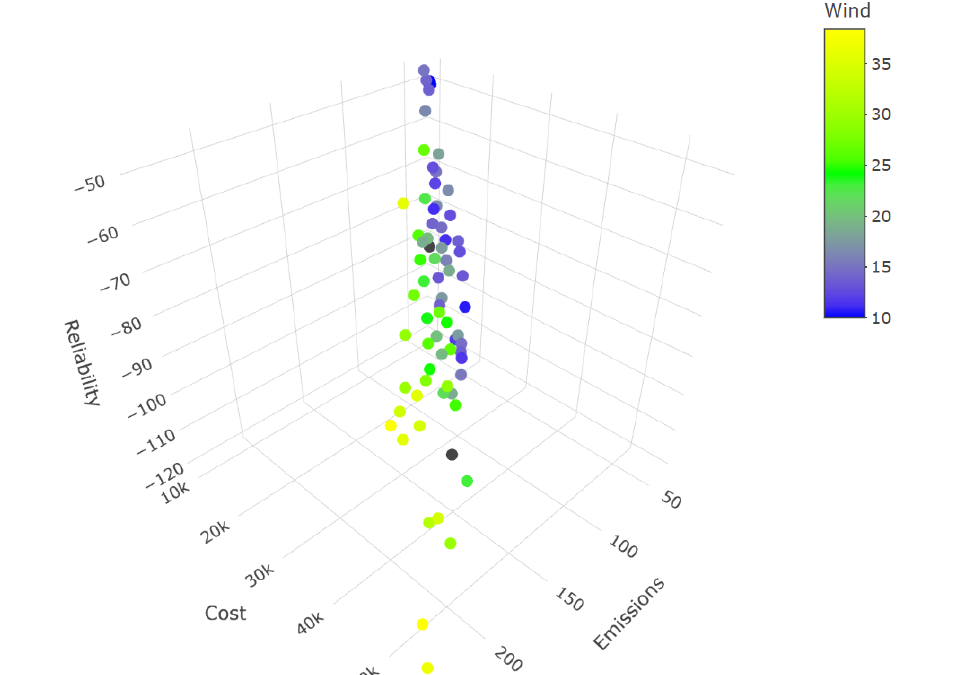} 
  			
  		}
  		
  		\caption[3D Pareto Front of Energy System Optimization (Colored by Wind Contribution)]{3D Pareto Front of Energy System Optimization (Colored by Wind Contribution).}\label{fig:fig-wind-contribution}
  	\end{figure}
  \end{Schunk}
  
  \begin{Schunk}
  	\begin{figure}[htbp]
  		
  		{\centering \includegraphics[width=4in]{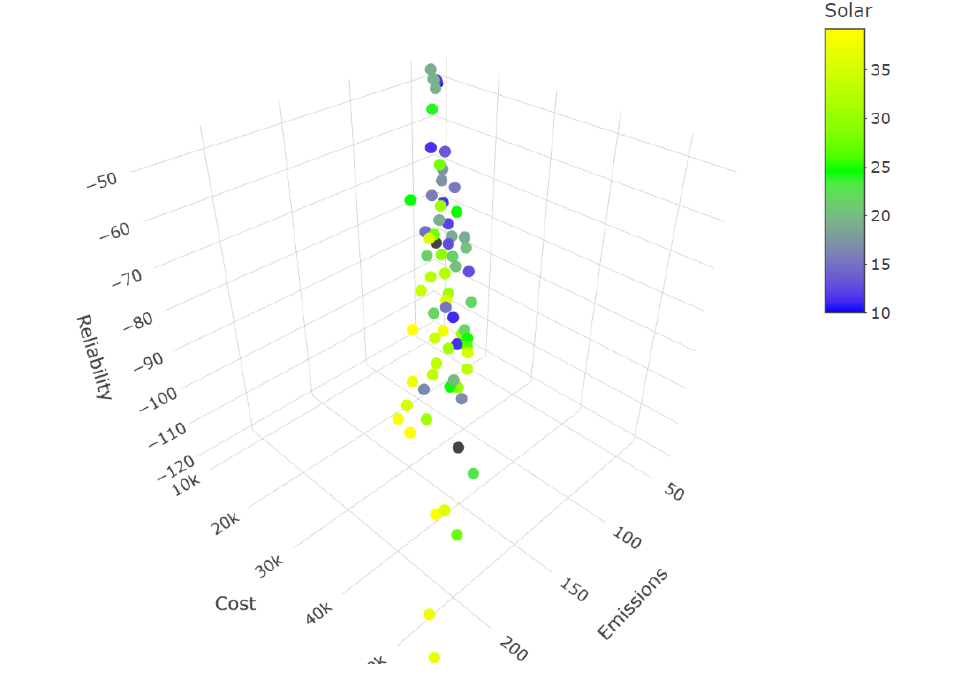} 
  			
  		}
  		
  		\caption[3D Pareto Front of Energy System Optimization (Colored by Solar Contribution)]{3D Pareto Front of Energy System Optimization (Colored by Solar Contribution).}\label{fig:fig-solar-contribution}
  	\end{figure}
  \end{Schunk}
  
  \begin{Schunk}
  	\begin{figure}[htbp]
  		
  		{\centering \includegraphics[width=4in]{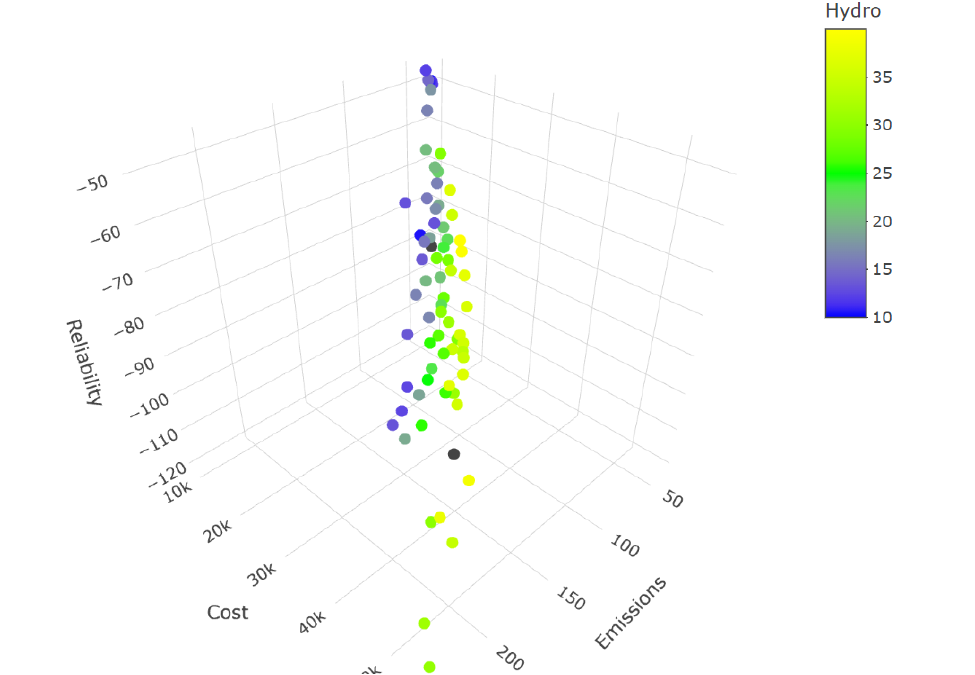} 
  			
  		}
  		
  		\caption[3D Pareto Front of Energy System Optimization (Colored by Hydro Contribution)]{3D Pareto Front of Energy System Optimization (Colored by Hydro Contribution).}\label{fig:fig-hydro-contribution}
  	\end{figure}
  \end{Schunk}
  
  The results underline the potential of the \textbf{Jaya} optimizer for
  addressing multi-objective optimization challenges in the energy domain.
  The visualizations provide decision-makers with a comprehensive view of
  the trade-offs among key objectives, enabling informed decisions based
  on policy priorities or operational needs.
  
  For instance, solutions emphasizing wind and solar contributions align
  with emissions reduction targets, while hydro and storage contributions
  are better suited for enhancing reliability. The flexibility to
  incorporate domain-specific constraints and the ability to explore a
  diverse set of trade-offs without requiring parameter tuning underscore
  the robustness of the \textbf{Jaya} package. These attributes make it a
  powerful tool for optimizing energy systems under real-world conditions,
  particularly in the context of renewable energy integration and
  sustainability planning.
  
  \hypertarget{summary}{%
  	\subsection{Summary}\label{summary}}
  
  The \texttt{Jaya} R package provides an efficient and flexible
  implementation of the parameter-free Jaya optimization algorithm for
  single-objective and multi-objective problems. This paper highlights the
  key features of the package, including constraint handling, adaptive
  population adjustment, Pareto front tracking, early stopping, and
  parallel processing. The \texttt{jaya()} function addresses
  single-objective optimization needs, while the \texttt{jaya\_multi()}
  function extends the algorithm for multi-objective scenarios,
  identifying non-dominated solutions for Pareto-optimal trade-offs. The
  package also includes robust helper functions, such as
  \texttt{summary.jaya()} for summarizing results and \texttt{plot.jaya()}
  and \texttt{plot\_jaya\_multi\_pairwise()} for visualizing optimization
  performance and Pareto fronts. These features make the package suitable
  for a wide range of applications, from engineering to operational
  research. A practical case study on energy optimization demonstrates the
  utility of the package, showcasing its capability to balance conflicting
  objectives like cost, carbon emissions, and system reliability in
  renewable energy integration. The results underline the adaptability and
  computational efficiency of the Jaya R package, making it a valuable
  tool for researchers and practitioners alike.
  
  \bibliography{RJreferences.bib}
  
  \address{%
  	Neeraj Dhanraj Bokde\\
  	Renewable and Sustainable Energy Research Center,\\%
  	Technology Innovation Institute, Abu Dhabi, 9639,\\ United Arab
  	Emirates\\
  	\textit{ORCiD: \href{https://orcid.org/0000-0002-3493-9302}{0000-0002-3493-9302}}\\%
  	\href{mailto:neeraj.bokde@tii.ae}{\nolinkurl{neeraj.bokde@tii.ae}}%
  }
  
\end{article}

\end{document}